\documentclass[twocolumn,prl,english]{revtex4-1}
\usepackage[T1]{fontenc}
\usepackage[latin9]{inputenc}
\usepackage{verbatim}
\usepackage{amsmath}
\usepackage{amssymb}
\usepackage{graphicx}
\usepackage{babel}

\makeatletter

\usepackage{xcolor}
\newcommand*\diff{\mathop{}\!\mathrm{d}}

\begin{document}

\title{Motor-free contractility in active gels}

\author{Sihan Chen$^{1,2}$, Tomer Markovich$^2$, Fred C. MacKintosh$^{1,2,3,4}$}
\affiliation{$^1$Department of Physics and Astronomy, Rice University, Houston, TX 77005\\
$^2$Center for Theoretical Biological Physics, Rice University, Houston, TX 77005\\
$^3$Department of Chemical and Biomolecular Engineering, Rice University, Houston, TX 77005\\
$^4$Department of Chemistry, Rice University, Houston, TX 77005}

\begin{abstract}
	Animal cells form contractile structures to promote various functions, from cell motility to cell division.
	Force generation in these structures is often due to molecular motors such as myosin that require polar substrates for their function.
	Here, we propose a motor-free mechanism that can generate contraction in biopolymer networks without the need for polarity.
	This mechanism is based on {\it active} binding/unbinding of crosslinkers that breaks the \emph{principle of detailed balance}, together with the asymmetric force-extension response of semiflexible biopolymers.
	We find that these two ingredients can generate steady state contraction via a non-thermal, ratchet-like process.	
	We calculate the resulting force-velocity relation using both  {coarse-grained} and microscopic models.
\end{abstract}
\maketitle

In living cells, most force generation is due to molecular motors that move in a directed manner, such as linearly along a substrate. In animal cells, for instance, myosin motors moving along polar actin filaments can drive large scale contraction and force generation~\cite{Alberts,	Fletcher2010485}. 
While this is especially apparent in muscle cells with ordered arrays of actin and myosin, a similar mechanism is at play in a wide range of cellular processes in non-muscle cells, including cell migration, establishment of cell polarity and even developmental processes at the multi-cellular level~\cite{Heisenberg2013948}.
At the molecular scale, such motor activity fundamentally depends on a combination of broken time and spatial symmetries. Motors such as myosin undergo a directed cycle of transitions among conformational states~\cite{Spudich2001387} that violates the principle of \emph{detailed balance} (DB) ~\cite{Boltzmann1872,Julicher1997, Battle2016604,Nardini2017,Markovich2020}. 
Such a cyclical reaction manifests a directionality in time and is only possible in steady-state with the consumption and dissipation of energy, {\it e.g.}, in the turnover of adenosine triphosphate (ATP) or other metabolic components~\cite{Alberts}. 
But, even an energy consuming reaction is insufficient to generate directed motion or force. 
For this, a spatial symmetry must also be broken. This is usually due to the polarity of the substrate such as actin, with well-defined \emph{plus} and \emph{minus} ends, to which a motor such as myosin couples. The resulting \emph{rectification} of the motor's conformational transitions to generate linear motion and do useful work is reminiscent of the Smoluchowsky-Feynman thermal ratchet~\cite{Feynman1963,Magnasco1993,Prost1994,Julicher1997}. 

Here, we propose a mechanism for motor-free contractility that depends on non-equilibrium activity but without polarity or broken spatial symmetry of the substrate. 
The fundamentally contractile nature of this mechanism manifests the broken time-reversal symmetry and violation of DB. 
But, contractility can occur even with apolar filaments such as intermediate filaments (IF) or in unpolarized arrays or disordered networks of filaments. 
As we show, contractility is a natural consequence of the asymmetric force-extension of semiflexible filaments such as actin or IFs in the cytoskeleton, which effectively rectifies the stochastic binding and unbinding and breaks the spatial symmetry, with the direction of motion always being in the `contractile' direction.
This can happen, provided that the transient (un)binding is active, {\it e.g.}, depends on metabolic components such as ATP. This model may provide a basis for understanding recently reports of myosin-independent contractility in cells~\cite{Wloka2013, Xue2016}. 

 {\textit{Basic mechanism} \textendash}
In our  {coarse-grained} model, we consider a substrate such as an unpolarized, disordered network of filaments, to which a particular semiflexible filament can bind. 
For simplicity, we treat each segment between crosslinks as a filament that can bind (unbind) to (from) a continuum viscoelastic substrate [Fig.~1(a)]. 
In the binding state, the polymer exerts tension on the gel, $\tau(\ell)= dU_e(\ell)/d\ell$, resulting in an average contractile force
\begin{equation}
\begin{aligned}
\langle F_s \rangle_\ell=\int P_{\rm on}(\ell)\tau(\ell)\diff\ell \, .
\end{aligned}
\label{e1}
\end{equation}
Here $\ell$ is the end-to-end length of the polymer, $U_e$ is the potential of mean-force (PMF) of the polymer, 
and $P_{\rm on}$ is the probability  for the polymer to have a given end-to-end distance $\ell$ in the bound state. 
If the (un)binding is passive, then $P_{\rm on}(\ell)$ is governed by a Boltzmann distribution,  DB is satisfied and the contractile force vanishes.
However, when the binding and/or unbinding process are active, {\it e.g.}, due to consumption or catalysis of a metabolic component such as ATP,   
$P_{\rm on}$ is generally not a Boltzmann distribution.
This, together with an asymmetric PMF, can lead to a net contractile force.
\begin{figure*}[ht]
	\centering
	\includegraphics[scale=0.425]{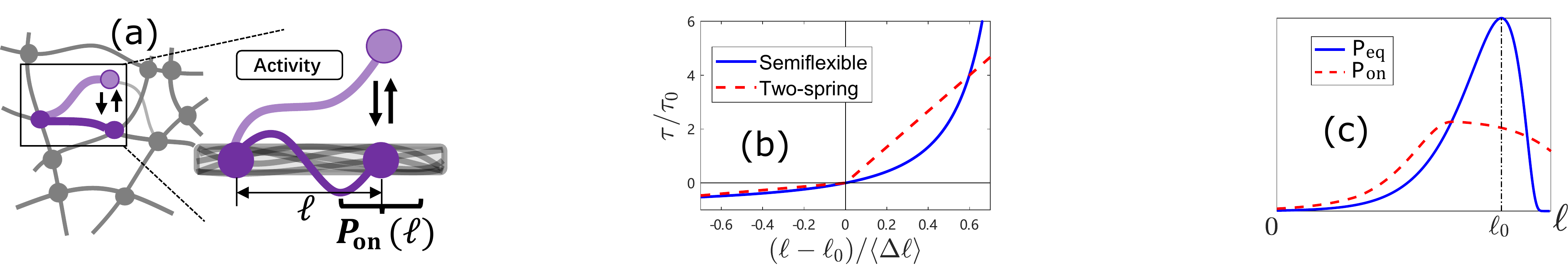}
	\caption{(a) Illustration of the  {coarse-grained} model. One polymer segment from a disordered network undergo binding/unbinding process attaining a steady-state length distribution, $P_{\rm on}$. 	
		The bound and unbound states are denoted by dark and light purple, respectively. 
		(b) Force-extension relation for inextensible semiflexible polymer ($\tau_0/\mu \to 0$ in Eq.~(\ref{e2})) and the two-spring potential with $K_1 = 2\tau_0/3\langle\Delta\ell\rangle$ and $K_2=10K_1$. 
		(c) Schematic plot (blue line) of the equilibrium distribution of a semiflexible polymer end-to-end length, and a plot of $P_{\rm on}$ (red dashed line) for a rigid substrate with $d = 4.2\langle\Delta \ell\rangle$ and $\omega_{\rm on}\gg \omega_{\rm off}$.
	}
	\label{Fig.0}
\end{figure*}

The force-extension relation for a semi-flexible polymer is [Fig.~\ref{Fig.0}(b)]~\cite{MacKintosh1995,Odijk1995,Wilhelm1996,Storm2005191,Broedersz2014}: 
	\begin{equation}
	\begin{aligned}
	\frac{\ell(\tau)-\ell_0}{\langle\Delta\ell\rangle} = \frac{\ell_0 }{\langle\Delta\ell\rangle} \frac{\tau} {\mu} + \epsilon\left[ \frac{\tau}{\tau_0} \left(1+\frac{\tau}{\mu}\right)\right] 
	\, , 
	\end{aligned}
	\label{e2}
	\end{equation}
where 
$\tau_0 \equiv \pi^2 k_B T  /  (6\langle\Delta\ell\rangle)$ is a characteristic tension, 
$\langle\Delta\ell\rangle \simeq \ell_0^2/(6\ell_p)$ is the mean end-to-end thermal contraction of a stiff polymer of rest length $\ell_0$ and persistence length $\ell_p$.
Here, $\mu$ is the enthalpic stretch modulus, $k_B$ is the Boltzmann constant, $T$ is the temperature and $\epsilon(\phi) = 1-3\frac{\pi \sqrt\phi \coth(\pi \sqrt\phi )-1} {\pi^2\phi}$~\cite{Broedersz2014}. 
Equation~(\ref{e2}) shows strong asymmetry: the polymers are hard to stretch and easy to compress.
Although this particular form of the force-extension relation is valid for contour lengths less than $\ell_p$, as is appropriate for cytoskeletal filaments, a qualitatively similar asymmetric relation holds in the opposite limit of semiflexibility, such as for DNA with small persistence length~\cite{Marko1994}.  A similar asymmetry may arise from the crosslinkers themselves (e.g., filamin), although this is not considered here ~\cite{Gardel20061762,Broedersz2008}.
As we show, such an asymmetry is sufficient to generate contraction, even for a symmetric substrate binding potential.
To demonstrate the sufficiency of this condition, we first consider a simplified and analytically tractable asymmetric  \emph{two-spring} force-extension relation
\begin{equation}
\begin{aligned}
\tau(\ell)=\left\{ 
\begin{aligned}
K_1(\ell-\ell_0) \qquad(\ell<\ell_0)\\
K_2(\ell-\ell_0) \qquad(\ell\geq \ell_0)
\end{aligned}
\right.
\end{aligned} \, ,
\label{e3}
\end{equation}
where  $K_1 < K_2$ are different spring constants for compression and stretching, respectively.

In the following, we consider the substrate to be isotropic with a characteristic spacing $d$ between binding sites. 
The binding rate can then be written as $\omega_{\rm on}P_b(\ell_b)$, where $\omega_{\rm on}^{-1}$ defines the average time spent in the unbound state and $P_b(\ell_b)$ is the probability to bind at an initial end-to-end polymer length $\ell_b$.
We divide the  {binding/unbinding} process into four steps:
(i) in the unbound state we assume the polymer quickly relaxes to an equilibrium distribution with rest length $\ell_0$ (Fig.~\ref{Fig.0}(c)), 
(ii) the polymer binds to the substrate at rate $\omega_{\rm on}$ and initial end-to-end length $\ell_b$ with probability $P_b(\ell_b)$, 
(iii) once the polymer binds, it contracts the gel due to its PMF, $U_e$, 
and (iv) the polymer {\it actively} unbind at constant rate $\omega_{\rm off}$. 
Here, we neglect thermal aspects of unbinding that we assume to be dominated by active processes. 

 {\textit{Limits of small and large fluctuations} \textendash}
In our model, in addition to $d$, another key length scale is the width, $\delta\ell$, of the thermal distribution of the filament end-to-end distance, $\ell$.  
We begin by considering a simplified limit in which this width is narrow compared with $d$, {\it i.e.}, $\delta\ell\ll d$.
We also assume a symmetric binding potential for which common motor activity is inhibited, although our mechanism does not require this symmetry. Together with the network isotropy this implies that the binding probability, $P_b(\ell_b)$, is symmetric around $\ell_b=\ell_0$ when $\delta \ell \ll d$~\cite{Supp}. 
We consider the simplest symmetric form for this probability distribution, characterized by just a width $d$ that represents the typical spacing of binding sites:
\begin{equation}
\begin{aligned}
P_b(\ell_b)=\frac{1}{d} \quad \mbox{for}\quad \ell_0-\frac{d}2 <\ell_b<\ell_0+\frac d 2\, ,
\end{aligned}
\label{e4}
\end{equation}
and $P_b=0$ otherwise. 

\begin{figure*}[ht]
	\centering
	\includegraphics[scale=0.55]{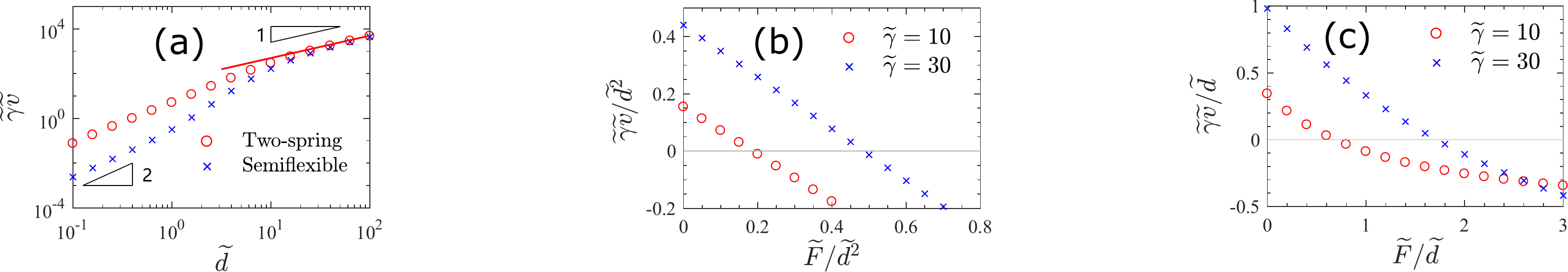}
	\caption{
		Numerical results for the contractile velocity.
		(a) Contractile velocity as function of the typical binding-site spacing for $\omega_{\rm off}\rightarrow \infty$. 
		For the semiflexible PMF, we use 
		$\mu = 4.37\times 10^{-8} \rm N$, $\tau_0 = 0.68 {\rm pN}$, $\ell_0 = 1\rm \mu m$, $\delta\ell= 6.2 {\rm nm}$, 
		while for the \emph{two-spring} PMF we choose $K_2 = \mu/\ell_0$ and $K_1 = 0.35\tau_0/\delta \ell$ (both PMFs have the same $\delta \ell$). 		
		The solid red line is the large-$d$ analytical solution of Eq.~(\ref{e7}).
		In (b) and (c) we plot the force-velocity relation for the two-spring and the semiflexible PMFs, respectively.
		The dimensionless quantities used are: $\tilde d = d/\delta \ell$, $\tilde\gamma = \gamma\omega_{\rm off} \delta \ell /\tau_0$, 
		$\tilde v = v/ (\omega_{\rm off}\delta \ell)$ and $\tilde F = F/\tau_0$. 
		Parameters used: (b) $\tilde d = 0.1$ , and (c) $\tilde d = 10$.  
		Both $\tilde F$ and $\tilde \gamma \tilde v$ are further rescaled according to their $d$ dependence (quadratic for large $\tilde d$ and linear for small $\tilde d$).
		In (a), (b) and (c) $\omega_{\rm on}\gg\omega_{\rm off}$. 	
	}
	\label{Fig.1}
\end{figure*}
 {For constant on and off rates, the survival probability for the polymer length $\ell$ in the bound state is $P_s(\ell;\ell_b,t)=\delta(\ell-\ell_f)e^{-\omega_{\rm off}t}$, where $\ell_f(\ell_b;t)$ is the trajectory of the polymer length given that the polymer binds at $t=0$ with length $\ell_b$ (thermal noise is neglected). Averaging this survival probability over time and all possible $\ell_b$, gives the steady-state length distribution $P_{\rm on}(\ell)=C_{\rm on}\int \diff \ell_b P_b(\ell_b)\int \diff t \omega_{\rm off}P_s(\ell;\ell_b,t)$. Here $C_{\rm on}=\omega_{\rm on}/(\omega_{\rm on}+\omega_{\rm off})$ is the fraction of time spent in the bound state.
For a rigid substrate, $\ell_f(\ell_b;t)=\ell_b $, and $P_{\rm on}(\ell)=C_{\rm on}P_b(\ell_b=\ell)$, leading to a contractile force $\langle F_s \rangle _{\ell} = dC_{\rm on}(K_2-K_1)/8$.}
In fact, any symmetric distribution $P_b$ 
generates a net contractile force.
For a compliant substrate such as a viscoelastic gel that resists deformation, this force will drive contraction. 
We consider two limiting behaviors of such a gel: a viscous liquid and an elastic solid.

For a viscous substrate response, with an effective drag coefficient $\gamma$,
$\ell_f$ follows:
\begin{equation}
\begin{aligned}
\gamma \frac{\diff{\ell}_f(\ell_b;t)}{\diff t} = -\tau\left(\ell_f\right)+F \, ,
\end{aligned}
\label{e6}
\end{equation}
where $F$ is an external load, and the average contractile velocity is given by $v=\left[\langle F_s \rangle_\ell-F\right]/\gamma$.
The contractile velocity for the simple two spring PMF (Eq.~\ref{e3}) with vanishing load is~\cite{Supp},
\begin{equation}
\begin{aligned}
v=&\frac{d\gamma C_{\rm on}\omega_{\rm off}^2(K_2-K_1)}{8(K_1+\gamma \omega_{\rm off})(K_2+\gamma \omega_{\rm off})} \,.
\end{aligned}
\label{e7}
\end{equation} 
The analytical expression for a finite load appears in~\cite{Supp}.
In the elastic limit, $\langle F_s\rangle_\ell$ has the same form as Eq.~(\ref{e7}) with $\gamma \omega_{\rm off} \to K_s$, an effective spring constant \cite{Supp}. 

For  {the $\delta\ell\gg d$ limit,} one can calculate the binding probability distribution $P_b(\ell_b)$ perturbatively, since it reduces to the 
equilibrium end-to-end distribution $P_{\mbox{ riptsize eq}}$ (Fig.~\ref{Fig.0}(c)) in the limit $d\rightarrow0$. 
For an analytic force-extension relation $\tau(\ell)$, a possible asymmetry will appear to lowest order as an anharmonic second derivative $\tau''$. 
Thus, the expected scaling of the net/average force from Eq.~(\ref{e1}) is given by $\langle F_s \rangle\sim\tau'' d^2$, leading to $v\sim d^2$~\cite{Supp}.
Therefore, a crossover from a quadratic to linear dependence on $d$ is expected. 

 {\textit{Force-velocity relations} \textendash}
To find the velocity for general $d$, we developed a numerical simulation
that takes into account both thermal fluctuations in $P_b(\ell_b)$ and a thermal Langevin noise in Eq.~(\ref{e6}) \cite{Supp}.
The length in the unbound state is then sampled from its equilibrium distribution $P_{\rm{ eq}}$ (Fig.~\ref{Fig.0}(c)), 
and $P_b$ of Eq.~(\ref{e4}) is assumed to be distributed symmetrically about this sampled length, over a width of order $d$. Such a binding probability results in an equal chance of the polymer being  stretched or compressed when it binds, which effectively flattens $P_{\rm eq}$ and leads to a non-Boltzmann $P_{\rm on}$ (Fig.~\ref{Fig.0}(c)).
The contractile velocity for the two-spring (Eq.~(\ref{e3})) and the semiflexible polymer (Eq.~(\ref{e2})) PMFs
are similar in both the small and large $d$ limits, see Fig.~\ref{Fig.1}(a).  
For $d \ll \delta\ell$, our numerical simulation verifies the expected $v \sim d^2$ for both PMFs~\cite{Supp}. 
For large $d$, due to the sampling of the extremes in the semiflexible PMF, the stretch modulus $\mu$ begins to dominate and 
the system becomes well-described by the two-spring potential with $K_2\rightarrow \mu/\ell_0$ and $K_1 \rightarrow 0$, which represents a \emph{rope-like} limit~\cite{Supp}.
Here, we observe both the predicted linear dependence of Eq.~(\ref{e7}) (red line in Fig.~\ref{Fig.1}(a)), as well as the expected coincidence of the two models for the parameters chosen. 
Our results for the two-spring potential do not depend qualitatively on $K_1$ or $K_2$ as long as these spring constants are not equal. Importantly, our model suggests that contractility is enhanced for rigid polymers, in contrast with motor-driven contractility in apolar arrays of filaments that is suppressed by filament rigidity \cite{Silva2011,Lenz2012}.

In Fig.~\ref{Fig.1}(b)-(c) we plot force-velocity curves for the small and large $d$ limits, respectively. 
In Fig.~\ref{Fig.1}(b) we use the two-spring potential, while in Fig.~\ref{Fig.1}(c) we use the semiflexble polymer PMF. 
In both cases, the velocity is reduced by an applied load, in a way similar to molecular motors~\cite{Derenyi19966775}. 
We find that decreasing $\tilde \gamma$ results in lower velocities and correspondingly lower stall forces.
This is due to the increased compliance and stress relaxation of the substrate, which lowers the substrate friction. 
\begin{figure*}[ht]
	\centering
	\includegraphics[scale=0.38]{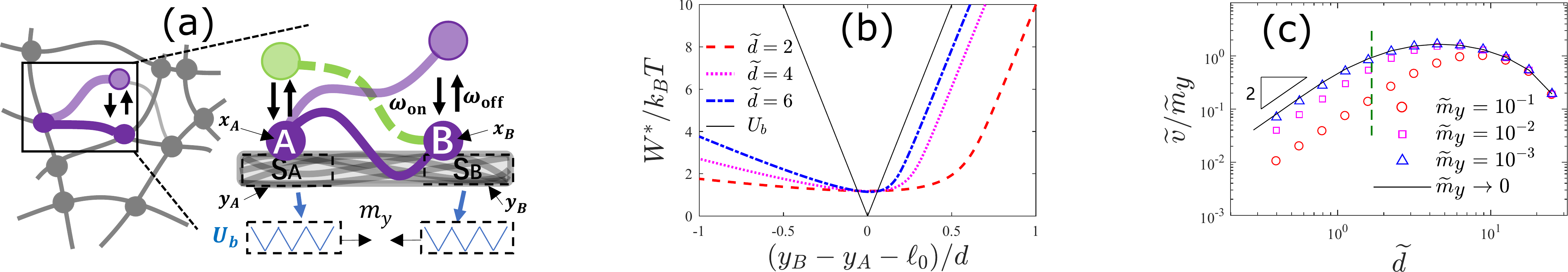}
	\caption{		
		(a) Illustration for the microscopic model. 
		 {Two crosslinkers actively bind/unbind to two different substrate regions. The binding potential $U_b$ for the two crosslinkers is assumed to be periodic with average spacing $d$ between binding sites. After binding the two regions are being pulled towards each other by the polymer contraction.}
		(b) Asymmetric effective  {interaction} $W^*$ (see~\cite{Supp} Eq.~(S28)) for various values of rescaled binding site size, $\tilde d = d/\delta \ell$, 
		using the same semiflexible potential as Fig.~2(a). 
		(c) Rescaled contractile velocity vs $\tilde d$ for various values of $\tilde m_y$ for $m_x \gg m_y$. 
		Here $\tilde v = v/(2\omega_{\rm off}\delta \ell) $, $\tilde m_y = m_y \tau_0 /(2\omega_{\rm off}\delta \ell)$ and $\omega_{\rm on}\gg \omega_{\rm off}$.
		Dashed green line indicates a binding site size of $d=10\rm nm$ and $\delta \ell = 6.2\rm nm$, as appropriate for a $1\mu{\rm m}$ actin filament.
	}
	\label{Fig.4}
\end{figure*}

 {\textit{Microscopic justification} \textendash} Having demonstrated the conceptual mechanism for contraction  {with a coarse-grained model, 
we now introduce a microscopic model for binding/unbinding of active crosslinks that illustrates in greater detail how the broken spatial symmetry arises even for an initially symmetric binding potential.
As we show, however, this microscopic model reduces to the coarse-grained model
in the appropriate limits, thus providing a microscopic justification for the simplified model above.}
As sketched in Fig.~\ref{Fig.4}(a), 
we consider a semiflexible segment whose two ends, $\rm A$ and $\rm B$, 
bind and unbind to  {two regions on a viscous substrate, $\rm S_A$ and $\rm S_B$, with constant rates $\omega_{\rm on}$ and $\omega_{\rm off}$, hence breaking DB}. 
 {The substrate regions  are assumed to be rigid and are able to diffuse independently. The interaction between these regions is only considered through their motion within the viscous network.}  {We describe the effective binding potential $U_b$ with a periodic triangular potential of depth $\Delta E$ and period $d$. As appropriate for a substrate consisting of apolar filaments or a disordered network, each well within $U_b$ is symmetric, while periodicity is not essential.} This binding potential may be measured by methods similar to previous ligand-receptor binding experiments \cite{Florin1994415}.
The PMF of the polymer is $U_e(x_B-x_A)$, where $x_A$ and $x_B$ are the positions of $\rm A$ and $\rm B$, respectively. 

Consider a binding event in which one of the polymer ends  {($\rm A$ by assumption)} is bound to the substrate at $t<0$ while the other binds at $t=0$.
 {The positions of $\rm S_A$ and $\rm S_B$ are described by the positions of the nearest binding sites to $\rm A$ and $\rm B$ at $t=0$, $y_A$ and $y_B$, respectively. After binding, the interaction between the polymer ends ($U_e(x_A-x_B)$) and the interactions of the polymer ends with the two substrate regions ($U_b(x_A-y_A)$ and $U_b(x_B-y_B)$) change the distance between the two substrate regions, $y_B-y_A$, which also describes the substrate length. This continues until either of the polymer ends unbinds at $t=t_e$, thus ending the bound state. In each binding event the substrate length change is $\Delta y = \langle y_B-y_A \rangle_{t=0} - \langle y_B-y_A \rangle_{t=t_e}$.} The averages are taken with respect to ${\cal P}(x_A,x_B,y_A,y_B;t)$, the survival probability for the positions of the polymer ends and the substrate regions, which follows a standard four-variable Fokker-Planck equation~\cite{Brackley2017,Supp}
with potential $W(x_A,x_B,y_A,y_B)=U_e(x_B-x_A)+U_b(x_A-y_A)+U_b(x_B-y_B)$, 
mobilities $m_x$ and $m_y$ (for both polymer ends and for both regions, respectively), and unbinding rate  {$2\omega_{\rm off}$ (both ends can unbind).} The dynamics in the unbound state are assumed to be fast \footnote{relaxing these assumptions will not change our results, but will make the formalism more intricate.}, 
leading to an initial, Boltzmann distribution of the polymer end-to-end distance, governed by  {${\cal P}(x_A,x_B,y_A,y_B;t=0)\sim \exp [-[U_e(x_B-x_A)+U_b(x_A-y_A)]/k_B T]$. 
Finally, The average contractile velocity (over many binding/unbinding events) is
$v =  \Delta y/\cal T$, where ${\cal T}=1/(2C_{\rm on}^2\omega_{\rm off})$ is the average time between two binding events. Importantly, this microscopic model reduces to the coarse-grained model when $m_y\ll m_x$ and $\Delta E\gg k_B T,\mu d^2/\ell_0$~\cite{Supp}.}

To further elucidate how the PMF asymmetry leads to contraction, we consider the following limits: (i) $m_x \gg m_y$, a physical limit corresponding to a polymer network substrate, whose dynamics are slower than those of a single polymer, and (ii) $\tau_r\ll\omega_{\rm off}^{-1}\ll\tau_{\rm hop}$, where $\tau_r$ is the average relaxation time of the polymer end within the binding site and $\tau_{\rm hop}$ is the average time for the polymer end to hop to a nearby binding site. This limit is applicable for  $\Delta E\gg k_BT$~\cite{Supp}.  With these assumptions one may treat $x_A$ and $x_B$ as fast variables and only consider the motion within  their initial binding sites. The dynamics of the distance $y_B-y_A$ is then reduced to an effective 1D Fokker-Planck equation for ${\cal P}(y_B-y_A;t)$\footnote{To be precise, eliminating $x_A$ and $x_B$ results in a 2D Fokker-Planck equation for ${\cal P}(y_B-y_A,y_B+y_A;t)$, 
but it can be shown~\cite{Supp} that ${\cal P}(y_B-y_A,y_B+y_A;t) = {\cal P}(y_B-y_A;t) {\cal P}(y_B+y_A;t)$ and $y_B+y_A$ has diffusive dynamics}, with
an effective interaction between $\rm S_A$ and $\rm S_B$, $W^*(y_A-y_B)$, which is the averaged $W$ in the equilibrium distribution of $x_A$ and $x_B$~\cite{Magnasco19942656,Supp}. As shown in Fig.~3(b), $W^*$ becomes asymmetric for finite $d$, and its shape becomes more asymmetric for small $d$.
Interestingly, a particle moving in a periodic $W^*$ is mathematically equivalent to a motor binding on a polar filament~\cite{Julicher1997}, 
with the crucial distinction that the motor directional motion is dictated by the filament polarity, while the motion here is always contractile in character. 

After obtaining the effective potential, we use the effective 1-D Fokker-Planck equation to find $\Delta y$ and the contractile velocity numerically \cite{Supp}.
Figure 3(c) shows the dependence of the contractile velocity on $d$. 
When $d \ll \delta \ell$, the velocity has quadratic dependence on $d$.
However, since  $\Delta E$ is finite, for large enough $d$, the binding potential becomes flat.
The velocity is then decreased for $d$ exceeding a threshold that depends on $\Delta E$ and $ m_y$.
The non-monotonic $v$ implies that there is an optimal ${d}$. 

 {\it{Conclusions} \textendash} 
We have presented  {a coarse-grained model and a microscopic model} to demonstrate how contraction can result from active (un)binding that violates DB.  {Taking $d\simeq10$nm, of order the size of a globular protein or the spacing of binding sites on a cytoskeletal filament, and $\delta\ell\simeq6$nm, corresponding to an actin filament of length $1\mu$m, both models predict a maximum contractile force $\sim0.5 \rm pN$.}
The active process leads to a breaking of time-reversal symmetry, much as does the enzymatic cycle of molecular motors, {\it e.g.}, fueled by the hydrolysis of ATP. 
Unlike a molecular motor, however, we show that contractility in our model doesn't depend on polarity or broken spatial symmetry of either filament or substrate.
Rather, the mechanical asymmetry that is generic for any filamentous biopolymer is sufficient to \emph{rectify} the active, cyclical (un)binding and generate contractile motion. 
Thus, this mechanism can lead to motor-like contractility even for apolar filaments, such as \emph{intermediate filaments}, which has not been thought possible. For polar filaments, mechanical asymmetry and buckling have also been shown to facilitate contractility by molecular motors~\cite{Silva2011,Lenz2012}, which can be distinguished from the present mechanism by its opposite dependence on filament rigidity. 

In addition to cytoskeletal filaments, crosslinking proteins such as filamin may also provide the mechanical asymmetry needed for rectification and contractility. Interestingly, disordered/apolar actin bundles have been shown to exhibit contraction when crosslinked with filamin \cite{Weirich2017}, although the active, non-equilibrium aspect of this system is unclear.
Our model may provide an explanation for recent observations of non-myosin-dependent dynamics of the contractile ring during cell division or cellularization \cite{Wloka2013, Xue2016}.
It is possible that ATP-dependent crosslinking by myosin in disordered actin networks or other structures may generate contractility~\cite{Ramaswamy2004,Shen2006,Wang2011,Silva2011,Alvarado2013,Tjhung2013,Markovich2019}, 
without the need for a motor \emph{power stroke}. 
It is even possible that septins, which are known to play a role in the contractile ring, may be responsible for force generation, even though they form apolar filaments \cite{Mavrakis2014,Valadares2017}. In any case, our model suggests a generic, steady-state mechanism for contraction even in apolar or fully disordered structures.

\begin{acknowledgements}
\emph{Acknowledgements}:
This work was supported in part by the National
Science Foundation Division of Materials Research
(Grant No.\ DMR-1826623) and the National Science
Foundation Center for Theoretical Biological Physics
(Grant No.\ PHY-2019745). The authors
acknowledge helpful discussions with M.~Gardel and J.~Theriot.
\end{acknowledgements}

\bibliography{citation}

\end{document}


\title{Supplementary Material \\ Motor-free contractility in active gels}
\author{Sihan Chen$^{1,2}$, Tomer Markovich$^{2}$ and F.\ C.\ MacKintosh$^{1,2,3,4}$}
\affiliation{
	$^{1}$Department of Physics and Astronomy, Rice University, Houston, TX 77005 \\
	$^{2}$Center for Theoretical Biological Physics, Rice University, Houston, TX 77005 \\
    $^{3}$Department of Chemical and Biomolecular Engineering, Rice University, Houston, TX 77005 \\
    $^{4}$Department of Chemistry, Rice University, Houston, TX 77005}

\maketitle
	
\vspace{-1cm}



\section{Coarse-grained model}
\subsection{Validation of the symmetric binding probability assumption}

We give here the theoretical background for the expression of the binding rate as written after Eq.~(3) of the main text and validate the assumption of symmetric binding probablity as written in Eq.~(4) of the main text. In the  {coarse-grained} model, we consider binding/unbinding of a segment of semiflexible polymer to/from a substrate. 
Let us denote the binding and unbinding rates as $\Omega_{\rm on}(\ell_u,\ell_b)$ and $\Omega_{\rm off}(\ell_b,\ell_u)$,
where $\ell_b$ and $\ell_u$ are the polymer length in the bound and unbound state, respectively. 
For a system obeying detailed balance, the transition rates satisfy~\cite{Julicher1997},
\begin{equation}
\Omega_{\rm on}(\ell_u,\ell_b)/\Omega_{\rm off}(\ell_b,\ell_u)=e^{\beta [U_e(\ell_u)-U_e(\ell_b)-U_b]},
\label{e1}
\end{equation}
with $U_b$ being the binding energy and $U_e$ the elastic potential of mean-force (PMF). 
This relation results in a Boltzman steady state distribution.
In practice, $U_b$ is large compared to both the thermal energy and elastic PMF, $|U_b|\gg k_B T, U_e(\ell)$. 
Hence, the binding process is mostly determined by $U_b$, which can be regarded as a substrate property.
Because we assume the substrate is isotropic (and therefore has also translational symmetry), 
$\Omega_{\rm on} (\ell_u,\ell_b) = \Omega_{\rm on} (\ell_u-\ell_b)$.
Being isotropic, the substrate also has parity symmetry,
allowing us to write $\Omega_{\rm on}(\ell_u,\ell_b)=\omega_{\rm on}P_c(|\ell_b-\ell_u|)$, where $P_c(|\ell_b-\ell_u|)$ 
is the probability distribution of the length change in the binding process, and $\omega^{-1}_{\rm on}$ defines the average time spent in the unbound state.

We assume $P_c$ can be characterized by the average binding site separation, $d$.
For simplicity we also take the dynamics in the unbound state to be fast compared to the time-scales of binding/unbinding and the relaxation of the substrate, 
hence, the unbinding length probability is the equilibrium distribution, $P_{\rm eq}(\ell_u) \sim \exp\left[-U_e(\ell_u)/k_B T\right]$. 
The probability for the polymer to bind with length $\ell_b$ is then
 \begin{equation}
 \begin{aligned}
 P_{b}(\ell_b) = \int P_{\rm eq}(\ell_u) P_c(|\ell_b-\ell_u|)\diff \ell_u \, .
 \end{aligned}
 \label{e3}
 \end{equation}
%
When $d$ is much larger than the width of the distribution $P_{\rm eq}(\ell_u)$ ($d\gg \delta \ell $) 
one may write $P_{\rm eq}(\ell_u) \simeq \delta(\ell_u - \ell_0)$,
leading to a symmetric binding probability $P_b(\ell_b) = P_c(|\ell_b-\ell_0|)$.

\subsection{The width of the equilibrium distribution}

In this subsection we calculate the width of the equilibrium distribution for both the two-spring and the semiflexible PMFs. 
The distribution width is denoted by $\delta\ell$ and is defined as the square root of the distribution variance, $\delta \ell^2 = \langle \ell^2\rangle-\langle \ell\rangle^2$. 
The averages are taken with respect to the equilibrium distribution, $P_{\rm eq}(\ell)=\exp{[-U_e(\ell)/k_BT]}/Z$,
with $Z$ being the partition function, $Z =\int \exp{[-U_e(\ell)/k_BT]} \diff \ell$.
For the two-spring PMF (Eq.~(3) in the main text), $U_e$ is harmonic and $\delta\ell$ reads,
%
\begin{equation}
\delta \ell =\left[ \frac{(\pi-2)K_1^2+\pi K_1^{\frac 3 2}K_2^{1/2}+4	K_1 K_2+\pi K_1^{1/2}K_2^{3/2}
	+(\pi-2)K_2^2 }{\pi K_1K_2\left(K_1^{1/2}+K_2^{1/2}\right)^2}k_B T\right]^{1/2} \, .
\label{e31}
\end{equation}

For the semiflexible PMF, we only consider the (realistic) case in which $\mu\gg\tau_0$, {\it i.e.}, the polymer is nearly inextensible.
In that case, we can take $\delta\ell$ to be the one of an inextensible polymer. 
This leads to $\delta \ell = \,\ell_0^2/(\sqrt{90}\ell_p)$~\cite{Broedersz2014}.

\subsection{Large $d$ limit for the two-spring PMF}

Let us calculate the contractile velocity (force) for a viscous (elastic) substrate in the limit of large $d$ ({\it i.e.} $d \gg \delta \ell$) for the two-spring PMF (Eq.~(3) in the main text).
Our goal is to calculate $\langle F_s \rangle_\ell$ of Eq.~(1) in the main text:
 {
\begin{equation}
\begin{aligned}
\langle F_s \rangle_\ell=\int P_{\rm on}(\ell)\tau(\ell)\diff\ell \, .
\end{aligned}
\label{efs}
\end{equation}
}
To do so we need to find $P_{\rm on}$, which for constant on and off rates (as we consider in the main text) reads, 
 {
\begin{equation}
P_{\rm on}(\ell) = C_{\rm on} \int \diff \ell_b P_b(\ell_b) \int \diff t \omega_{\rm off}P_s(\ell;\ell_b,t) \, .
\label{ep1}
\end{equation}
Here $C_{\rm on}=\omega_{\rm on}/(\omega_{\rm on}+\omega_{\rm off})$ is the fraction of time spent in the bound state,
	and $P_s(\ell;\ell_b,t)=\delta(\ell-\ell_f)e^{-\omega_{\rm off}t}$~\cite{Phillips2009}, where $\ell_f(\ell_b;t)$ is the trajectory of the polymer length given that the polymer binds at $t=0$ with length $\ell_b$ (thermal noise is neglected). }
Below we detail the calculation of $P_{\rm on}$ and $\langle \tilde F_s \rangle_\ell$ for the various cases.

\subsubsection{Contractile velocity for a viscous substrate}

We continue by calculating the contractile velocity for the two-spring potential, which is plotted in Fig.~2 of the main text. For a viscous substrate, the polymer length in the bound state, $\ell_f$, is a dynamic variable obeying a Langevin equation (Eq.~(5) of the main text). Solving this equation gives the trajectory $\ell_f(\ell_b;t)$:
\begin{equation}
\ell_f(\ell_b;t)=
\left\{ 
\begin{aligned}
&\ell_0+\frac{F}{K_1}+\left(\ell_b-\ell_0-\frac{F}{K_1}\right)e^{-{K_1}t/\gamma }  \qquad&(\ell_b<\ell_0 \quad {\rm{and}} \sl\quad t<t_0)\\
&\ell_0+\frac{F}{K_2}\left(1-e^{-{K_2}(t-t_0)/\gamma }\right)  \qquad&(\ell_b<\ell_0 \quad {\rm and} \quad t\geq t_0)\\
&\ell_0+\frac{F}{K_2}+\left(\ell_b-\ell_0-\frac{F}{K_2}\right)e^{-{K_2}t/\gamma } \qquad&(\ell_b\geq \ell_0) \, ,
\end{aligned}
\right.
\label{e6}
\end{equation}
where $t_0 \equiv  \frac{\gamma}{K_1}{\rm ln}(\frac{\ell_0-\ell_b+F/K_1}{F/K_1})$ is the time for which $\ell_f(t_0)=\ell_0$. 
 {With this result we find explicitly the survival probability distribution of the polymer length $\ell$, 
$P_s(\ell,\ell_b;t)=\delta(\ell-\ell_f)\exp(-\omega_{\rm off}t)$, which is then used to calculate the contractile force using Eqs.~(\ref{efs}) and (\ref{ep1}).}
The contractile velocity under external force $F$ for a viscous substrate with viscosity $\gamma$ is defined as 
\begin{equation}
v = \left[\langle F_s \rangle_\ell-F\right]/ \gamma \, .
\label{ev}
\end{equation}
Using the probability distribution above with Eqs.~(\ref{efs})-(\ref{e6})  we obtain
\begin{equation}
\begin{aligned}
v&=\frac 1 \gamma \left[- F+\int \diff\ell P_{\rm on}(\ell)\tau(\ell) \right]\\
&=\bigg((K_2-K_1)\Big[(2F^2+dFK_1)\Big(1+\frac{K_1 d}{2F}\Big)^{-{\gamma \omega_{\rm off}}/{K_1}}-2F^2\Big]\\
&\qquad+(K_1-\gamma\omega_{\rm off})\Big[\frac{1}{4}d^2 \gamma\omega_{\rm off}(K_2-K_1)-dF(K_1+K_2+2\gamma \omega_{\rm off})\Big]\bigg)\\
&\qquad\qquad\times \frac{\omega_{\rm on}\omega_{\rm off}}{2d\gamma(\omega_{\rm on}+\omega_{\rm off})(K_1^2-\gamma^2\omega^2_{\rm off})(K_2+\gamma\omega_{\rm off})}-\frac{F\omega_{\rm off}}{\gamma(\omega_{\rm on}+\omega_{\rm off})}.
\end{aligned}
\label{e7}
\end{equation}
For vanishing external force, $F=0$, the contractile velocity is calculated by taking limit of $F\rightarrow 0$ in Eq.~(\ref{e7}), yielding Eq.~(6) of the main text.

\begin{figure}[t!]
	\includegraphics[scale=0.7]{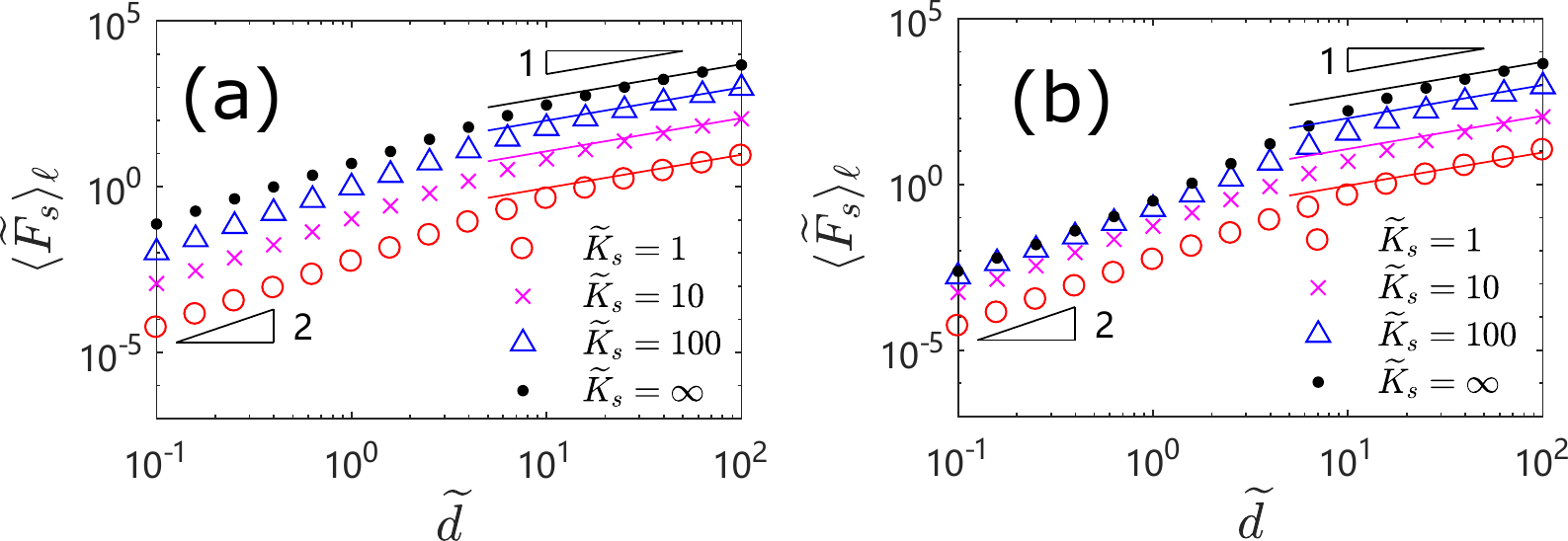}
	\caption{
		Contractile force as function of the typical binding-site spacing  using (a) the two-spring PMF and (b) the semiflexible PMF. 
		For the semiflexible PMF, we use 
		$\mu = 4.37\times 10^{-8} \rm N$, $\tau_0 = 0.68 {\rm pN}$, $\ell_0 = 1\rm \mu m$, and $\delta\ell= 6.2 {\rm nm}$, 
		while for the two-spring PMF $K_2 = \mu/\ell_0$ and $K_1 = 0.35\tau_0/\delta \ell$ (both PMFs have the same $\delta \ell$). 
		The rescaled quantities are: $\tilde d = d/\delta \ell$, $ \langle\tilde F_s \rangle_\ell = \langle F_s \rangle_\ell /\tau_0$ and $\tilde K_s = K_s \delta \ell/\tau_0$.   
		The solid lines are the large $d$ analytical solution of Eq.~(\ref{e11}). 
		In both figures we assume $\omega_{\rm on} \gg \omega_{\rm off}$.
	}
	\label{Fig.4}
\end{figure}
%

\subsubsection{Contractile force for an elastic substrate}

 {For an elastic substrate with spring constant $K_s$, the polymer length in the bound state, $\ell_f(\ell_b;t)$, is time-independent and can be written as $\ell_f(\ell_b)$, since the substrate reacts immediately  to an external force. The length in the bound state
is calculated using the force-balance equation, $\tau(\ell_f) = K_s \left(\ell_f-\ell_b\right)$, yielding,}
%
%
\begin{equation}
\begin{aligned}
\ell_f(\ell_b;t)=
\left\{ 
\begin{aligned}
\ell_0+(\ell_b-\ell_0)\frac{K_s}{K_1+K_s} &\qquad(\ell_b<\ell_0)\\
\ell_0+(\ell_b-\ell_0)\frac{K_s}{K_2+K_s} &\qquad(\ell_b\geq \ell_0)\,.
\end{aligned}
\right.
\end{aligned}
\label{e10}
\end{equation}
 {We then find the survival probability $P_s(\ell_f;\ell_b,t)$, and because $\ell_f$ is not a function of $t$ we have $\int \diff t \omega_{\rm off}\delta(\ell-\ell_f)e^{-\omega_{\rm off}t}=\delta(\ell-\ell_f)$. Equation (\ref{efs}) then reads:}
%
\begin{equation}
\begin{aligned}
\langle F_s\rangle_\ell =\int\diff\ell P_{\rm on}(\ell)\tau(\ell) = C_{\rm on} \int\diff\ell_b P_b(\ell_b)\tau\left[\ell_f(\ell_b)\right]  =\frac{C_{\rm on}K_s^2(K_2-K_1)}{(K_1+K_s)(K_2+K_s)}\frac d 8\, .
\end{aligned}
\label{e11}
\end{equation}
In Fig.~S1(a) we plot this contractile force as function of $d$. 
The numerical results in the large $d$ limit agree well with Eq.~(\ref{e11}). 
Since the force-extension relation is linear for both stretching and compression (for large $d$), the contractile force shows linear dependence on $d$. 
We also observe decreasing force for softer substrates, where the contractile force is reduced due to the deformation of the substrate. Note that the softness of the substrate is a collective property that is distinguished from the individual polymer rigidity as it also depends on polymer density, crosslinkers density etc. 
In the small $d$ regime we see that $\langle F_s \rangle \sim d^2$ which will be explained below in Section~F.

\subsection{Semiflexible PMF in the large $d$ limit}

In this subsection we demonstrate that the semiflexible PMF can be approximated using the ``two-spring''  PMF in the large $d$ limit, both of which are plotted in Fig.~2(a) of the main text. The force-extension relation of an extensible semiflexible biopolymer with rest length $\ell_0$, persistence length $\ell_p$ and stretch modulus $\mu$ is written in Eq.~(2) of the main text. 
The inverse relation $\tau(\ell)$, cannot generally be written explicitly, but as we show below, it can be approximated in the large $d$ limit.
In Eq.~(2) we use the function  $\epsilon(\phi) = 1-3\frac{\pi \sqrt\phi \coth(\pi \sqrt\phi )-1} {\pi^2\phi}$ 	to describe the force-extension relation of inextensible polymer~\cite{Broedersz2014}.  
This function has two asymptotic limits:
\begin{equation}
\begin{aligned}
\epsilon(\phi)=
\left\{ 
\begin{aligned}
&1 \qquad \quad \,\,\,\, (\phi\rightarrow \infty) \\
&-\infty \qquad (\phi\rightarrow -1) \, ,
\end{aligned}
\right.
\end{aligned}
\label{e13}
\end{equation}
and the corresponding limits for $\tau(\ell)$ are:
\begin{equation}
\begin{aligned}
\tau(\ell)\approx
\left\{ 
\begin{aligned}
&\frac{\mu}{\ell_0}(\ell-\ell_0) \qquad (\ell-\ell_0\gg\delta\ell)\\
&-\tau_0 \qquad \qquad \,\,(\ell_0-\ell\gg\delta\ell) \, .
\end{aligned}
\right.
\end{aligned}
\label{e14}
\end{equation}
Equation~(\ref{e14}) shows that the polymer behaves like a spring with spring constant $\mu/\ell_0$ under large extension,
while under compression it generates a constant force. 
It is then possible to approximate the restoring force by a ``two-spring'' potential that generates the same average force for both compression and extension.
This leads to a spring constant of $2\tau_0/d$ for compression. 
To conclude, in the large $d$ limit, the semiflexible PMF is well approximated  by a ``two-spring'' PMF  with $K_2 = \mu/\ell_0$ and $K_1\to0$ (the rope limit).
This is also verified in Fig.~2(a) of the main text.

\subsection{Details of the numerical simulation}

This section details the numerical procedure we use in order to calculate  the contractile velocity (force) for a viscous (elastic) substrate for a general $d$.
As we explain below, this includes accounting explicitly for fluctuations of both the polymer and the substrate.
The aim in both subsections below is to find $P_{\rm on }$ of Eq.~(\ref{ep1}) 
which is then substituted in  {Eq.~(\ref{efs})} to give $\langle F_s\rangle$, and 
the contractile velocity is obtained using Eq.~(\ref{ev}). 
%

\subsubsection{Contractile velocity for a viscous substrate}

For a viscous substrate, thermal fluctuations should be considered in both $P_b$ and the Langevin equation (Eq.~(5) of the main text).
The length after binding, $\ell_b$, is obtained using Eq.~(\ref{e3}), where $P_c$ is chosen to be a squared distribution with width $d$ (similarly to Eq.~(4) of the main text):
\begin{equation}
P_{c}(|\ell_b-\ell_u|) = \frac 1 d, \qquad(|\ell_b-\ell_u|<\frac d 2)
\label{e2}
\end{equation}
and $P_c=0$ otherwise. 
The resulting $P_b$ is then
 \begin{equation}
\begin{aligned}
P_{b}(\ell_b) = \frac{1}{d}\int_{\ell_b-d/2}^{\ell_b+d/2} P_{\rm eq}(\ell_u) \diff \ell_u \, ,
\end{aligned}
\label{e32}
\end{equation}
where $P_{\rm eq}(\ell)=\exp(-U_e(\ell)/k_B T)/\int \diff \ell \exp(-U_e(\ell)/k_B T) $ is the equilibrium distribution in the unbound state. 
The thermal noise in the Langevin equation modifies Eq.~(5) of the main text:
\begin{equation}
\begin{aligned}
\gamma \dot{\ell}_f = -\tau(\ell_f)+F +\eta(t)\, ,
\end{aligned}
\label{e15}
\end{equation}
where $\eta(t)$ is a Gaussian white noise with zero mean and variance, $\langle \eta(t)\eta(t')\rangle = 2 k_B T \gamma\delta (t-t')$. 
 {This Langevin equation is equivalent to the following Fokker-Planck equation for the survival probability $P_s(\ell;\ell_b,t)$~\cite{Gardiner}:
\begin{equation}
\begin{aligned}
\frac{\partial}{\partial t} {P_s}(\ell_f;\ell_b,t) - \frac{1}{\gamma} \frac{\partial}{\partial \ell_f}  \left[(U_e+F\ell_f){P_s}+k_B T\frac{\partial}{\partial \ell_f} P_s \right] = -\omega_{\rm off}{P_s}(\ell_f;\ell_b,t) \,,
\end{aligned}
\label{e16}
\end{equation}
 {where the right-hand-side is the result of the constant off rate $\omega_{\rm off}$.}
The initial condition is  $P_s(\ell_f;\ell_b,t=0) = \delta(\ell_f-\ell_b)$.} We calculate  {$P_s(\ell_f;\ell_b,t)$} by numerically solving Eq.~(\ref{e16}). 
Finally we use this result to calculate $P_{\rm on }$ of Eq.~(\ref{ep1})  {and $\langle \tilde F_s \rangle_\ell$} of Eq.~(\ref{efs}). 

\subsubsection{Contractile force for an elastic substrate}

For an elastic substrate thermal fluctuation are accounted in both $P_b$ and the state of the substrate (it is fluctuating about its rest position),
where the polymer length after binding, $\ell_b$, is the same as in the viscous substrate case, Eq.~(\ref{e32}).  {To describe the state of the elastic substrate, we denote the extension of the substrate length about its rest length by $s$ ($s=0$ means the substrate does not generates force). The elastic energy generated by the substrate is then $K_s s^2/2$. }
%
The extension just after binding, $s_b$, is sampled from the equilibrium distribution of the substrate,  $P_{s_b}$,
which is a Gaussian distribution with zero mean ($\langle s_b \rangle = 0$) and variance $\langle s^2_b \rangle = K_BT / K_s$. 

 {We continue by calculating the equilibrium distribution of the polymer length $\ell_f$ in the bound state, $P_f(\ell_f;\ell_b,s_b)$, given that at $t=0$ the polymer length is $\ell_b$ and the substrate extension is $s_b$.}
Since the polymer is bound to the substrate, its two ends do not move relative to the substrate,  { thus imposing a constraint $\ell_f -\ell_b = s - s_b$.} 
The equilibrium distribution of $\ell_f$ for a given $\ell_b$ and $s_b$ is: 
\begin{equation}
\begin{aligned}
P_f(\ell_f;\ell_b,s_b)=\frac 1 {Z(\ell_b,s_b)} \exp\left[{-[U_e(\ell_f)+K_s(\ell_f-\ell_b+s_b)^2/2]/k_B T}\right] \, ,
\end{aligned}
\label{e17}
\end{equation}
where $Z =\int\diff \ell \, \exp\left({-[U_e(\ell)+K_s(\ell-\ell_b+s_b)^2/2]/k_B T}\right)  $ is the partition function. 
 {The survival probability distribution of the polymer length is now found by choosing $s_b$ from its distribution, $P_{sb}$,
\begin{equation}
\begin{aligned}
P_s(\ell;\ell_b,t) = e^{-\omega_{\rm off}t}\int P_f(\ell;\ell_b,s_b)P_{s_b}(s_b)\diff s_b .
\end{aligned}
\label{e33}
\end{equation}
Equation~(\ref{e33}) is numerically calculated and used to find $P_{\rm on}$ (Eq.~(\ref{ep1})) for each $K_s$.}

\subsection{The small $d$ limit}
In this subsection we show analytically the quadratic dependence of the contractile velocity in the small $d$ limit, which is shown in Fig.~2(a) of the main text. When $d\to0$ the binding probability reduces to the equilibrium distribution, $P_{\rm eq}$.
For $d\ll \delta\ell$, the integration range in Eq.~(\ref{e32}) is within a small region around $\ell_u =\ell_b$, allowing us to expand $P_{\rm eq}(\ell_u)=\exp(-U_e(\ell_u)/k_BT)/[\int\exp(-U_e/k_BT) \diff\ell_u]$ around $\ell_b$:
\begin{equation}
\begin{aligned}
P_b(\ell_b)\simeq
\int_{\ell_b-d/2}^{\ell_b+d/2}\Bigg[1+(\ell_u-\ell_b)P'_{\rm eq}(\ell_b)+\frac{(\ell_u-\ell_b)^2}{2}P''_{\rm eq}(\ell_b)\Bigg]\diff \ell_u
= P_{\rm eq}(\ell_b)\left[1+\frac{d^2}{24}\left(\frac{U_e'^2(\ell_b)}{(k_BT)^2}-\frac{U_e''(\ell_b)}{k_B T}\right)\right]\,,
\end{aligned}
\label{e18}
\end{equation}
 {where $f'(\ell_b)=[\diff f(\ell) /\diff \ell]|_{\ell=\ell_b}$ and $f''(\ell_b)=[\diff^2 f(\ell) /\diff \ell^2]|_{\ell=\ell_b}$.}
As expected, Eq.~(\ref{e18}) shows that the equilibrium distribution is slightly perturbed by $d$, where the difference scales as $d^2$. 
The resulting steady-state distribution, $P_{\rm on}(\ell)$ (Eq.~(\ref{ep1})),
is also perturbed around the equilibrium distribution, and the deviation scales as $d^2$. 
Thus, we have $\langle F_s \rangle_\ell \sim d^2$, and both the contractile velocity for a viscous substrate and the contractile force for an elastic substrate show quadratic dependences on $d$. 

To show this explicitly, let us consider a nearly rigid substrate, {\it i.e.}, $ \gamma \omega_{\rm off} \rightarrow \infty$ or $ K_s\rightarrow \infty$. 
In this case the polymer is not relaxing when in the bound state ({\it i.e.} $P_s(\ell;\ell_b,t) \simeq \delta(\ell-\ell_b)\exp(-\omega_{\rm off}t)$), and produces an average contractile force (on the substrate) of
\begin{equation}
\begin{aligned}
\langle F_s \rangle_\ell =\int\diff\ell P_{\rm on}(\ell)\tau(\ell)=C_{\rm on}\int\diff\ell_b P_{b}(\ell_b)U_e'(\ell_b) =\frac{C_{\rm on}d^2}{24}\int \diff \ell_b P_{\rm eq}(\ell_b)U_e'''(\ell_b) \, ,
\end{aligned}
\label{e19}
\end{equation}
where we have used integration by parts. This force is positive for any potential with a positive $U_e'''(\ell)$. 
Note that the above result is obtained for constant on/off rates. 
In case that the on/off rates obey detailed balance (see Eq.~(\ref{e1})) $P_{\rm on}$ would remain the equilibrium distribution even for finite $d$ and $\langle F_s \rangle_\ell$ would vanish.

\section{Microscopic Model}

\subsection{4D Fokker-Planck equation}

Here we present the 4D Fokker-Planck equation that is used to describe the evolution of ${\cal P}(x_A,x_B,y_A,y_B,t)$ in the microscopic model. The microscopic model considered in the main text describes the motion of the two polymer ends ($\rm A$, $\rm B$) and the corresponding two  {proximal substrate regions} ($\rm S_A$, $\rm S_B$)
after a binding event that starts at $t=0$.  {The positions of $\rm A$ and $\rm B$ are denoted by $x_A$ and $x_B$, respectively. The substrate regions are assumed to be rigid and are able to diffuse independently. The interaction between these regions is only considered through their motion within the viscous network. The positions of $\rm S_A$ and $\rm S_B$ are described by the positions of the nearest binding sites to $A$ and $B$ at $t=0$, $y_A$ and $y_B$, respectively. When $t<0$, the polymer end $\rm A$ is bound while $\rm B$ is unbound (this is equivalent to the case in which $\rm B$ is bound while $\rm A$ is unbound).} $\rm B$ is attached to the substrate at $t=0$ and 
the survival probability of the positions of these four coordinates for $t>0$, ${{\cal P}(x_A,x_B,y_A,y_B;t)}$, 
can be described using a standard four-variable Fokker-Planck equation~\cite{Brackley2017}:

\begin{equation}
\begin{aligned}
&\partial_t {\cal P}(x_A,x_B,y_A,y_B;t)+\nabla \cdot {\bm {J}(x_A,x_B,y_A,y_B;t)}= - 2\omega_{\rm off}(x_B-y_B) {\cal P}\,,\\
&{\cal P}(x_A,x_B,y_A,y_B;t=0)=\frac{\chi(x_A,x_B,y_A,y_B)}{ Z}\exp{\Big[- [U_e(x_B-x_A)-U_b(x_A-y_A)]/k_B T\Big]}\,,
\end{aligned}
\label{e20}
\end{equation}
where $J_\alpha=-m_\alpha( k_B T\partial_\alpha {\cal P}+{\cal P}  \partial_\alpha W)$ without the summation convention.
We define $m_x$ and $m_y$ to be the mobilities of the polymer ends and  {the substrate regions}, 
thus $m_\alpha=m_x$ for $\alpha=x_A, x_B$ and  $m_\alpha=m_y$ for $\alpha=y_A, y_B$.
 {Here $\chi(x_A,x_B,y_A,y_B) = \Theta\left(d/2 - \left|x_A - y_A\right|\right) \Theta\left(d/2 - \left|x_B - y_B\right|\right)$ gives the  boundaries of the initial condition ($\Theta(x)$ is the Heaviside function), which appears since we define $y_A$ and $y_B$ to be the positions of the nearest binding sites of $\rm A$ and $\rm B$},
$W(x_A,x_B,y_A,y_B)=U_e(x_B-x_A)+U_b(x_B-y_B)+U_b(x_A-y_A)$ is the total energy in the bound state, and $Z$ is the partition function. 
The initial condition is chosen to be the equilibrium distribution in the unbound state. Here $U_e (x_B-x_A)$ and $U_b (x_A-y_A)$ stand for the elastic PMF and the effective binding potential for $\rm A$. Since in the microscopic model the details of the binding potential are accounted for, the binding itself does not change the positions of $\rm A$ and $\rm B$. Because $\rm B$ is not attached to the substrate at $t<0$, $y_B$ does not contribute to the initial condition.  {Furthermore, as $y_B$ is the position of the nearest binding site to $\rm B$, the distance between $x_B$ and $y_B$ must be smaller than $d/2$, and $y_B$ is sampled from a uniform distribution between $(x_B-d/2,x_B+d/2)$, because there is no energy associated with its position in the unbound state.}
We choose for simplicity the effective binding potential to be a periodic triangular potential with depth $\Delta E$ and period $d$, 
{\it i.e.}, $U_b(x) =  \frac{2\Delta E}{d} |x-nd|$ for $d(2n-1)/2 \leq x \leq d(2n+1)/2$, although periodicity is not essential. 

 {In order to find the contractile velocity, one needs to calculate the average distance between two binding sites at the end of the bound state,  
\begin{equation}
\begin{aligned}
\langle y_B-y_A\rangle _{t=t_e}=2\omega_{\rm off}\int_0^{\infty}\diff t\, (y_B-y_A)\,{\cal P}(x_A,x_B,y_A,y_B;t)\,,
\end{aligned}
\label{e21}
\end{equation}
where $t_e$ is the time in which one of the polymer ends unbind.}
The average distance at the beginning of the bound state is $\langle y_B-y_A\rangle _{t=0} = \ell_0$, which can be understood as follows.
The binding site position, $y_B$, is sampled uniformly around $B$, hence, $\langle y_B-x_B\rangle _{t=0}=0$.
 {Because the potential is symmetric we also have $\langle y_A-x_A\rangle _{t=0}=0$,
and since the polymer is relaxed in the unbound state with rest length $\ell_0$, $\langle x_B-x_A\rangle _{t=0}=\ell_0$. 
Then, $\Delta y = \langle y_B-y_A \rangle_{t=0} - \ell_0$.} 

	The contractile velocity is calculated using $v =  \Delta y/ {\cal T}$, where ${\cal T}$ is the average time between two binding events, which is the sum of the average lifetimes in the bound and unbound states. The average lifetime of the polymer in the bound state (both ends are bound) is $\tau_{\rm off}=1/2\omega_{\rm off}$, where the factor of 2 is due to the fact that both ends can unbind. The average lifetime in the unbound state is more complicated as it is composed of two different states: (i) both ends are unbound, and (ii) only one end is unbound. The fraction of time (with respect to the total time in both the bound and unbound states) spent in these two states can be written as, $C_1 = (1-C_{\rm on})^2$, $C_2 = 2C_{\rm on}(1-C_{\rm on})$, where $C_{\rm on}=\omega_{\rm on}/(\omega_{\rm on}+\omega_{\rm off})$ is defined in the main text as the fraction of time in which one of the polymer ends is bound to the substrate (regardless of the other end situation). Therefore, when the polymer is in the bound state, the probabilities to be in state (i) and (ii) are, $P_1=C_1/(C_1+C_2)$ and $P_2=C_2/(C_1+C_2)$, respectively. Since the unbound state can only be ended when the system is in state (ii), the net binding rate, given the polymer is in the unbound state, is $\omega^*_{\rm on}=P_2\omega_{\rm on}$, and the average lifetime of the unbound state is $\tau_{\rm on} = 1/\omega^*_{\rm on}$. Taken together, we have ${\cal T} = \tau_{\rm off}+\tau_{\rm on}=1/(2C_{\rm on}^2 \omega_{\rm off})$.

\subsection{Reducing to the coarse-grained model}

In the main text page 4 we claim that the microscopic model reduces to the coarse-grained model under certain conditions. This subsection is devoted to prove this claim. If $\Delta E/d\gg|\diff U_e(x_B-x_A)/\diff (x_B-x_A)|$, the binding potential is very steep thus imposing a large 
force on the polymer end with amplitude $2\Delta E/d$ (unless the polymer end is in the center of the potential well). 
Thus, the polymer end $B$ is dragged to the center of $S_B$ in a short time denoted as $\sigma$. 
For $0<t<\sigma$, we can deterministically estimate the motion of $B$ after it binds to the substrate:
\begin{equation}
\begin{aligned}
&\frac{\diff{x_B}}{\diff t} = \frac{2m_x\Delta E}d\, {\rm{sign}}(y_B-x_B)\\
&\frac{\diff{y_B}}{\diff t} = -\frac{2m_y\Delta E} d \,{\rm{sign}}(y_B-x_B)\,.
\end{aligned}
\label{e34}
\end{equation}
From Eq.~(\ref{e34}) we have $[x_B(t=\sigma)-x_B(t=0)]/m_x = -[y_B(t=\sigma)-y_B(t=0)]/m_y$, and since $x_B(t=\sigma) = y_B(t=\sigma)$, we can write,
\begin{equation}
\begin{aligned}
x_B(t=\sigma)-x_B(t=0) = \frac{m_x}{m_x+m_y} \left[(y_B(t=0)-x_B(t=0)\right]\,.
\end{aligned}
\label{e35}
\end{equation}
Finally, because $y_B(t=0)-x_B(t=0)$ has a uniform distribution of width $d$, 
the  polymer length distribution at $t=\sigma$ is the same as $P_b$ of Eq.~(4) of the main text but with spacing $d \to m_x d/(m_x+m_y)$.  {When $m_x \gg m_y$ we get the binding probability of the coarse-grained model. We now consider the other requirement, $\Delta E/d\gg|\diff U_e(x_B-x_A)/\diff (x_B-x_A)|$. If $d\leq\delta \ell$, we have $|\diff U_e(x_B-x_A)/\diff (x_B-x_A)|\sim k_B T/\delta \ell$, and $\Delta E\gg k_B T$ satisfies the requirement. If $d>\delta \ell$, we have $|\diff U_e(x_B-x_A)/\diff (x_B-x_A)|\sim \mu d/\ell$, and thus $\Delta E \gg \mu d^2 /\ell_0$. To conclude, in the limits $\Delta E\gg k_B T, \mu d^2/\ell_0 $ and $m_x\ll m_y$ the microscopic model reduces to the coarse-grained one.}

\subsection{Estimation of $\tau_r$ and $\tau_{\rm hop}$}
In this subsection we derive the expressions of $\tau_r$ and $\tau_{\rm hop}$ that are used in page 4 of the main text. 
$\tau_r$ is the time required for the polymer end to relax within one binding site. Let the binding site be within $(-d/2,d/2)$ and consider the diffusion of a particle with mobility $m_x$ within a triangular binding potential. We use the ``intrawell relaxation time'' introduced in Ref.~\cite{Derenyi1999} to estimate $\tau_r$. It is defined as the average mean-first-passage time of the particle from any fixed initial position $x_0$, to a final position $x$ that is sampled from a Boltzmann distribution governed by $U_b$:
\begin{equation}
\begin{aligned}
\tau_r &= \frac{1}{k_B T Zm_x}\int_{-d/2}^{d/2}\diff x\int_{x}^{d/2}\diff y\int_{y}^{d/2}\diff z\exp\left[[-U_b(x)+U_b(y)-U_b(z)]/k_B T\right]\approx \frac{ d^2}{2k_B Tm_x}\left(\frac{k_B T}{\Delta E}\right)\,,
\end{aligned}
\label{e30}
\end{equation}
where $Z = \int_{-d/2}^{d/2}\diff x \exp[-U_b(x)]$ is the partition function. In Eq.~(\ref{e30}), the integrals over $\diff y$ and $\diff z$ calculate the mean-first-passage time from $x_0$ to $x$, and the integral over $\diff x$ calculates the average mean-first-passage time with the Boltzmann weight of $U_b(x)$. Note that $\tau_r $ is independent of the initial position $x_0$.

$\tau_{\rm hop}$ is the average time for the polymer end to hop to another binding site. The time required to escape from a potential well can be estimated by the mean-first passage time from the bottom of the well to the top of the well, which is~\cite{Gardiner}:
\begin{equation}
\begin{aligned}
\tau_{\rm hop}&=\frac{1}{k_B Tm_x} \int_0^{d/2}\diff y\int_{-d/2}^y\diff z	\exp\left[[U_b(y)-U_b(z)]/k_B T\right]\approx \frac{d^2}{2k_B Tm_x}\exp\left(\frac{\Delta E}{k_B T}\right)\,.
\end{aligned}
\label{e31}
\end{equation}
We then conclude that for  $\Delta E\gg k_B T$, we have $\tau_r\ll\tau_{\rm off}\ll\tau_{\rm hop}$. It can be understood intuitively: in the large $\Delta E$ limit, the potential well is steep enough thus driving a fast relaxation within the potential well, while the high potential barrier prevents hopping towards another binding site. 

\subsection{Variable Elimination: From 4D Fokker-Planck	 Equation to a 1D Fokker-Planck Equation}

Here we simplify the 4D Fokker-Planck equation and obtain a  1D  Fokker-Planck equation for the distance between the two substrate regions. We further derive the effective potential $W^*$ that is plotted in Fig. 3(b) of the main text. As explained in the main text, we treat $x_A$ and $x_B$ as fast variables because the polymer ends mobility is much larger than that of the two substrate regions, $m_x \gg m_y$.
Together with $\tau_r\ll \tau_{\rm off}$, this allows us to write~\cite{Magnasco1994}
\begin{equation}
\begin{aligned}
{\cal P}(x_A,x_B,y_A,y_B;t)=\frac{\exp{[- W(x_A,x_B,y_A,y_B)/k_B T]}}{Z_Y(y_A,y_B)}{\cal P}_Y(y_A,y_B;t)\,,
\end{aligned}
\label{e25}
\end{equation}
with $Z_Y=\int \diff x_A \diff  x_B \chi \, {e^{- W/k_BT}}$. The appearance of $\chi$ (defined after Eq. (\ref{e20})) restrict the motion of the polymer end to be within one binding site, as appropriate for $\tau_{\rm off}\ll\tau_{\rm hop}$.
Because $\omega_{\rm off}$ is constant we can rewrite Eq.~(\ref{e20}) in a simple form:
\begin{equation}
\begin{aligned}
&\partial_t {\cal P}_Y(y_A,y_B;t)+\nabla \cdot{ \bm{J^Y}(y_A,y_B;t)}= - 2\omega_{\rm off} {\cal P}_Y(y_A,y_B;t)\,,\\
&{\cal P}_Y(y_A,y_B;t=0)=\int \diff x_A \diff x_B \,{\cal P}(x_A,x_B,y_A,y_B;t=0)\,,
\end{aligned}
\label{e26}
\end{equation}
where $J^Y_\alpha(y_A,y_B;t)=-m_y[{\cal P}_Y(y_A,y_B;t)\partial_\alpha W^*(y_A,y_B)+k_B T\partial_\alpha{\cal P}_Y(y_A,y_B;t)]$ ($\alpha = y_A, y_B$ without the summation convention). 
This is a 2D Fokker Planck equation with an effective 2D potential
\begin{equation}
\begin{aligned}
W^*(y_A,y_B)=\int \diff x_A \diff x_B \frac{\chi(x_A,x_B,y_A,y_B) \, e^{- W/k_BT}}{Z_Y(y_A,y_B)} W(x_A,x_B,y_A,y_B)\,.
\end{aligned}
\label{e27}
\end{equation}
Note that, although we use a periodic $U_b$, this periodicity is lost in $W^*$ due to $\chi$. This is because the polymer end cannot hop to a different binding site as explained in the subsection above.
Since the system is symmetric under translations, $W^*(y_A,y_B)$ can only be a function of $y_B-y_A$. 
Then, we perform substitution of variables, $u = y_B-y_A$, $r = y_B+y_A$, and as $W^*$ only depends on $u$, these two variables are decoupled. 
We find that $r$ follows simple diffusion dynamics, while $u$ can be described by a 1D Fokker-Planck equation with distribution $P_U$:
\begin{equation}
\begin{aligned}
&\partial_t {\cal P}_U(u;t)+\partial_u{ {J_U}(y_A,y_B;t)}= - 2\omega_{\rm off} {\cal P}_U(u;t)\,,\\
&{\cal P}_U(u;t=0)=\int \diff y_A  \,{\cal P}_Y(y_A,y_B=y_A+u;t=0)\,,
\end{aligned}
\label{e29}
\end{equation}
where ${J_U}=-2m_y[{\cal P}_U\partial_u W^*(u)+k_B T\partial_u{\cal P}_U]$. 
 {The average distance at the end of the bound state, Eq.~(\ref{e21}), is also modified:
\begin{equation}
\begin{aligned}
\langle y_B-y_A\rangle _{t=t_e}=2\omega_{\rm off}\int_0^{\infty}\diff t\, u \,{\cal P}_U(u;t)\,. 
\end{aligned}
\label{e28}
\end{equation}
}

\bibliography{citation}